\documentclass{birkjour}

\usepackage{graphicx}
\usepackage{amsmath}
\usepackage{amsfonts}
\usepackage{amssymb}
\usepackage{xcolor}
\usepackage{scalerel}
\usepackage{enumitem}
\usepackage{ulem}
\usepackage{multirow}
\usepackage{multicol}
\usepackage{MnSymbol}
\usepackage{stackrel}
\usepackage{bm}

\newcommand{\R}{\mathbb {R}}

\newcommand{\N}{\mathbb {N}}
\newcommand{\Z}{\mathbb {Z}}
\newcommand{\PP}{\mathcal{P}}
\newcommand{\QQ}{\mathcal{Q}}

\newcommand{\ann}{\operatorname{ann}}

\newtheorem{thm}{Theorem}[section]

\theoremstyle{definition}
\newtheorem{defn}[thm]{Definition}
\theoremstyle{remark}

\numberwithin{equation}{section}

\begin{document}

\title[Creative Telescoping on Multiple Sums] {Creative Telescoping on Multiple Sums}

\author[C. Koutschan]{Christoph Koutschan}

\address{%
RICAM, Austrian Academy of Sciences\\
Altenberger Stra\ss e 69\\
4040 Linz\\
Austria}

\email{christoph.koutschan@ricam.oeaw.ac.at}


\author[E. Wong]{Elaine Wong}

\address{%
RICAM, Austrian Academy of Sciences\\
Altenberger Stra\ss e 69\\
4040 Linz\\
Austria}

\email{elaine.wong@ricam.oeaw.ac.at}

\thanks{Both authors were supported by the Austrian Science Fund (FWF): F5011-N15.}



\keywords{Symbolic Summation, Creative Telescoping, Holonomic Function, Hypergeometric Series}

\date{March 17, 2021}

\begin{abstract}
We showcase a collection of practical strategies to deal with a problem arising from an analysis of integral estimators derived via quasi-Monte Carlo methods. The problem reduces to a triple binomial sum, thereby enabling us to open up the holonomic toolkit, which contains tools such as creative telescoping that can be used to deduce a recurrence satisfied by the sum. While applying these techniques, a host of issues arose that partly needed to be resolved by hand. In other words, no creative telescoping implementation currently exists that can resolve all these issues automatically. Thus, we felt the need to compile the different strategies we tried and the difficulties that we encountered along the way. In particular, we highlight the necessity of the certificate in these computations and how its complexity can greatly influence the computation time.
\end{abstract}

\maketitle

\section{The Problem}

Applications give an experiential realization to what is theoretically possible and illuminate computational issues that the theory may not have foreseen. In the following exposition, we describe a problem that was encountered in one mathematical area, that was resolved with the tools of another, but not without roadblocks. This story not only arrives at a nice conclusion (the problem could be solved), but along the way uncovers the subtleties of the theory that may not be immediately apparent to a new user. Who knows, these gems might also find itself to be useful in other contexts. It is with this goal in mind that this manuscript exists, and we hope that the reader will bear with us as we explicate the details and hint at where some improvements can be made.

Recently, Wiart and Wong~\cite{WiartWong20} derived a formula for the covariance of an integral estimator for functions satisfying a certain decay condition, based on a quasi-Monte Carlo framework developed by Wiart, Lemieux, and Dong~\cite{WiartLemieux19}. More specifically, the latter introduced some randomness into selected point sets to improve the uniform distribution of points, giving access to certain probabilistic error estimates. The former then proposed some conditions for functions in terms of their Walsh coefficients and they were able to deduce a simplified formula for the covariance, which measure the effectiveness of a randomized quasi-Monte Carlo (RQMC) estimator of the integral of the function over the unit hypercube using the randomized point set. The conclusion in the paper~\cite{WiartWong20} is that under the proposed conditions, the covariance can be shown to be not positive and therefore, the RQMC estimator performs better than the standard Monte Carlo estimator. This covariance formula is written as the following polynomial in~$x$,
\begin{equation}
\label{eq:mainpoly}
G_s(x):=\sum_{k=1}^{m+s-1}\Bigg(\sum_{r=1}^{s}\binom{s}{r}\binom{k-1}{r-1}\frac{b-1}{(-b)^r}\sum_{i=0}^{r-1-c_m(k)}(-b)^i\binom{r-1}{i}\Bigg)(bx)^k,
\end{equation}
where~$c_m(k)=\max(k-m,0).$ The goal in~\cite{WiartWong20} is to show that $G_s(x)\leq 0$ for all $b, m, s \in \N$, $b\geq 2$ and $x\in[0,1)$. Much of the QMC literature focuses on optimizing point sets to achieve desired distribution properties and using techniques in analysis to improve bounds on estimators (see \cite{DickHinrichsPillichshammer15} and \cite{DickPillichshammer10} for an overview). Unfortunately, we found that such techniques do not enable us to reach the desired conclusion. Accordingly, we choose to approach this problem by employing symbolic computation rather than analysis: using the available computer algebra software for holonomic functions~\cite{Kauers09,Koutschan10,Schneider07}, we carry out a guess-and-prove strategy that ultimately leads us to deduce a suitable closed form for~\eqref{eq:mainpoly}. The result is an expression in terms of regularized beta functions which allows us to show the desired nonpositivity statement.

This article serves to highlight the ``proving'' aspect of the strategy, i.e., the derivation and proof of a third-order recurrence that~\eqref{eq:mainpoly} satisfies. In computer algebra, such computations are typically handled by the method of \textit{creative telescoping} \cite{Zeilberger90}. Implementations already exist that serve this purpose \cite{Chyzak97,KauersJaroschekJohansson15,Koutschan09,Schneider07}, with the caveat that there are many strategies that could be used in combination to make the computation more effective, and in some cases resolve issues that the implementations don't do automatically. Thus, we implore the reader to not immediately jump into the first strategy that is presented here, but rather to think of it as a buffet, some of these might work really well for one problem, and not as much for another. We want to make it clear that a rigorous derivation and motivation for \eqref{eq:mainpoly} can already be found in \cite{WiartWong20} and we will not repeat what has already been written. Instead, our goals here are intended for two types of researchers:
\begin{itemize}
\item those who may want see the types of objects that are amenable to state-of-the-art symbolic computation techniques;
\item those who may be interested in automating symbolic computation software and are interested in seeing the grisly details and roadblocks that appear with the use of current methods.
\end{itemize}

\section{Background}\label{sec:background}

    On a first glance, we note that all constituents of~\eqref{eq:mainpoly} have the property of being ``holonomic''. For the purposes of this paper, we stay slightly informal (but still rigorously practical) and use the definition that holonomic functions are sequences which satisfy ``sufficiently many'' linear recurrences with polynomial coefficients. One convenience of dealing with such functions comes from the fact that holonomicity is preserved through basic operations: we will refer to these as ``closure properties''. For example, the product of two holonomic functions is again holonomic~\cite[Proposition 3.2]{Zeilberger90}. From a computational point of view, if we know the recurrences that the two holonomic functions satisfy, we can construct a recurrence that their product satisfies and even bound its order and the degrees of its polynomial coefficients.
    
    Our expression~\eqref{eq:mainpoly} consists of summation quantifiers, (products of) binomial coefficients, and polynomial/exponential functions in the parameters. The binomial coefficient, for example, can be described completely via two recurrences and finitely many initial conditions. From our usual notion of binomial coefficients, we can immediately write down the two recurrences (valid on $\mathbb{Z}\times\mathbb{Z}$):
    \begin{align}\label{eq:recbinom}
    \begin{split}
        (n-k+1)\binom{n+1}{k}-(n+1)\binom{n}{k}=0,\\
        (k+1)\binom{n}{k+1}-(n-k)\binom{n}{k}=0,
    \end{split}
    \end{align}
    and specify the initial conditions
    \[
      \binom{0}{0}=1,\quad \binom{-1}{0}=0,\quad \binom{-1}{-1}=0.
    \]
    By doing so, we interpret the binomial coefficient over the integers in the traditional (combinatorial) way, namely, it is nonzero only for $0\leq k\leq n$. We explicitly highlight this because the computer algebra system Mathematica uses a more general notion of the binomial coefficient, which extends its definition to the negative integers. It is important to note that this generalized binomial coefficient is defined by the very same recurrences, but just uses different initial conditions:
    \[
      \binom{0}{0}=1,\quad \binom{-1}{0}=1,\quad \binom{-1}{-1}=1.
    \]
    Unfortunately, the more general view of the binomial coefficient affects the natural boundaries of our summation limits: the summands containing these coefficients behave inappropriately outside of our prescribed bounds, which is of course irrelevant regarding the definition of~$G_s(x)$, but which may cause problems when evaluating boundary terms originating from the telescoping sums. We will have to address this issue later.
    
    In the following, we will use this simple bivariate sequence $\binom{n}{k}$ to illustrate the main features of the holonomic systems approach. Using the notation $S_n$ resp.\ $S_k$ to represent the forward shift operator in the given variable, we can rewrite \eqref{eq:recbinom} so that each of the corresponding operators
    \begin{align}\label{eq:bcann}
    \begin{split}
        (n-k+1)S_n-(n+1),\\
        (k+1)S_k-(n-k),
    \end{split}
    \end{align}
    maps $\binom{n}{k}$ to the zero sequence. We say that these operators annihilate the given function. As one can see, the translation between recurrence and operator can be read off immediately. Viewing recurrences as operators enables the use of algebraic methods to manipulate them more efficiently. However, we then have to deal with objects that do not always commute. The appropriate algebraic framework to represent such operators is an Ore algebra. In the following technical definition, $\partial$ serves as a placeholder for any of our operator symbols~$S_n$ or~$S_k$.
    
\begin{defn}\label{def:orealgebra}
  Let $R$ be a ring.
  \begin{enumerate}
    \item If $\sigma\colon R\to R$ is a ring endomorphism and $\delta\colon R\to R$ is such that addition and the ``skew'' Leibniz law is satisfied, that is, 
    \[\delta(f+g)=\delta(f)+\delta(g),\]
    \[\delta(fg)=\delta(f)g+\sigma(f)\delta(g),\]
    for all $f,g\in R$, then $\delta$ is called a \emph{$\sigma$-derivation.}
    
    \item Suppose now that there is an endomorphism $\sigma\colon R\to R$ and a $\sigma$-derivation $\delta\colon R\to R$. Suppose further that a ring structure is defined on the set $R[\partial]$ of univariate polynomials in~$\partial$ with coefficients in~$R$, equipped with the usual addition, and multiplication is such that
    \[
      \partial^i \partial^j=\partial^{i+j}
      \quad\text{and}\quad
      \partial f=\sigma(f)\partial+\delta(f)
      \qquad\text{ for all } i,j\in\N \text{ and } f\in R.
    \]
    Then $R[\partial]$ is an \emph{Ore algebra} over~$R$. We typically use the symbol~$\mathbb{O}$ to denote such algebras.
  
    \item Suppose that $f$ is in the left-$R[\partial]$-module $R$ with action $\cdot~\colon~\mathbb{O}\times R\rightarrow R$ such that $1\cdot f=f$ and $L_1\cdot(L_2\cdot f)=(L_1L_2)\cdot f$ for all $L_1,L_2\in \mathbb{O}$. Then we say
    \[
      \ann(f)=\{ L\in R[\partial] \mid L\cdot f = 0\}
    \]
    is the \emph{annihilator} of~$f$ in~$R[\partial]$. 
  \end{enumerate}
\end{defn}

    For the binomial coefficient $\binom{n}{k}$, the Ore algebra that we use is $\mathbb{O}=R[S_k,S_n]$ with $R=\mathbb{Q}(n,k)$. A quick check shows that shift operators satisfy the required commutation properties. In the definition of this Ore algebra, each $\sigma$ denotes a forward shift operation (clearly a ring endomorphism) and $\delta \equiv 0$ (clearly a $\sigma$-derivation).
    
    The reader may have wondered why we did not consider Pascal's rule as a potential defining recurrence for the binomial coefficient. The reason is that~\eqref{eq:recbinom} are quite canonical generators for the set of all recurrences satisfied by~$\binom{n}{k}$. In algebraic terms, we can formulate this statement precisely: the annihilator of~$\binom{n}{k}$, which is a left ideal in~$\mathbb{O}$, is generated by the operators~\eqref{eq:bcann}, let's call them $P_1,P_2\in\mathbb{O}$:
    \[
      \ann\binom{n}{k} = \{ C_1\cdot P_1 + C_2\cdot P_2 \;\mid\; C_1,C_2 \in\mathbb{O} \}.
    \]
    Moreover, the two operators in \eqref{eq:bcann} even form a (left) Gr\"obner basis of $\ann\binom{n}{k}$.
    
    The key tool used in rigorously deriving a ``grand'' recurrence for $G_s(x)$ lies in the highly touted creative telescoping algorithm~\cite{Zeilberger91} for symbolic sums and integrals, as implemented in the HolonomicFunctions package~\cite{Koutschan10}. In order to construct a recurrence for a symbolic parametric sum of the form
    \begin{equation}\label{eq:symbsum}
      \sum_k \mbox{summand},
    \end{equation}
    the algorithm takes as input a list of generators, like the ones in \eqref{eq:bcann}, for an annihilating ideal of the summand. If the summand is given as a closed-form expression, then such a list is automatically computed, provided that it is recognized to be holonomic. The algorithm then identifies lists of operators $\PP$ and $\QQ$ (in the form of Ore polynomials in the algebra as described in Definition~\ref{def:orealgebra}) such that for each $P\in\PP$ and its corresponding $Q\in\QQ$, the operator $P-(S_k-1)\cdot Q$ is an element of the given annihilating ideal. The set $\PP$ contains the so-called ``telescopers'' (all of which are free of $k$ and $S_k$ but may contain the other parameters), and the set $\QQ$ the corresponding ``certificates''.
    
    How do these objects help us? Summing with respect to~$k$ gives relations of the form
    \begin{equation}\label{eq:ct}
        \sum_k P\cdot \mbox{summand}-\sum_k(S_k-1)\cdot Q \cdot \mbox{summand} = 0.
    \end{equation}
    In a best-case scenario, each of the $P$'s commutes with the first summation in~\eqref{eq:ct} (allowing us to pull it out of the sum so that we can view the elements of $\PP$ being applied to the whole sum and not just the summand) and the second summation collapses to zero by telescoping (leaving no trace of the certificate). From there, we would conclude that $\PP$ generates a left ideal of annihilating operators for~\eqref{eq:symbsum}, that is, it represents a set of recurrences which are satisfied by the sum. Coming back to our triple sum~\eqref{eq:mainpoly}, we can repeatedly apply this process until a recurrence for the outermost (and hence the whole) sum is deduced.
    
    However, life is not always that easy: during the application of this strategy to the particular summation problem~\eqref{eq:mainpoly}, we encountered the following difficulties that are somewhat prototypical for the holonomic systems approach. This explains why, despite being automatable in principle, it still lacks a press-the-button implementation that would provide a computer proof of a claimed identity in a completely automatic way and without any human interaction.
    
    \begin{enumerate}[itemsep=2pt,parsep=2pt]

        \item The summand, i.e., the expression inside a summation quantifier, may take on nonzero values outside of the respective summation bounds. Thus, there is no reason to expect a~priori that the second summation in~\eqref{eq:ct} will evaluate to zero. And indeed, we found that it did not, and such terms constitute some of the ``inhomogeneous parts'' of the equation. An additional annihilator for them is required in order to homogenize the recurrence.
    
        \item The upper boundaries contain the variable~$s$, and the operators in~$\PP$ contain shifts in~$s$, causing difficulties with moving~$P\in\PP$ to be outside of the sum.
    
        \item Some of the operators in~$\QQ$ contain singularities at the boundary values so we were forced to exclude these values (which required compensation elsewhere). This is because the sum in~\eqref{eq:ct} containing the certificates is designed to collapse to only boundary value evaluations and we encounter problems if the summands are undefined at such values. Further issues could surface if those summands were also undefined at some intermediary value. Luckily, this was not the case here.
    
        \item Mathematica, in its symbolic zeal, rewrites the innermost sum in~\eqref{eq:mainpoly} as a hypergeometric $_2F_1$ series and the second innermost sum as a DifferenceRoot. While the values of the $_2F_1$ function match with our sum within the domain in question, there are still an infinite number of values for which it doesn't. The DifferenceRoot is Mathematica's version of a recurrence together with initial values, but unfortunately not helpful for our purposes because it is incompatible with HolonomicFunctions and does not support multivariate recurrences that are needed for creative telescoping.
    
    \end{enumerate}
    
    We illustrate the first three difficulties with a toy example. For thoroughness, we apply our strategy fully to this example to give a sufficient idea of the broader behavior. From this point on, we will periodically perform the service of demonstrating how computers and humans interact, by highlighting (in brackets) when paper-and-pencil reasoning is used and when automation is applied.
    
    Suppose we want to rigorously determine an annihilating operator for
    \begin{equation}\label{eq:binomialsum}
    \sum\limits_{k=5}^{n}\binom{n}{k}
    \end{equation}
    for $n\geq 5$. In other words, we would like to identify a recurrence that it satisfies. The creative telescoping algorithm (computer) outputs the telescoper $P=S_n-2$ and the certificate $Q=\frac{k}{k-n-1}$. Then \eqref{eq:ct} implies
    \[
    \sum\limits_{k=5}^{n}(S_n-2)\binom{n}{k}-\underbrace{\left(\frac{k}{k-n-1}\binom{n}{k}\Biggr|_{k=n+1}-\frac{k}{k-n-1}\binom{n}{k}\Biggr|_{k=5}\right)}_{\text{collapsed sum with singularity at } k=n+1}=0,
    \]
    with the summation containing the certificate collapsing to only evaluations at the boundary values. We note that $\binom{n}{k}$ is nonzero for $k=0,\ldots,4$, i.e., outside of the summation bounds. After substituting $k=5$ we get a nonzero contribution in the certificate (compare this to the situation where the lower bound is $\leq 0$). We next note that the certificate has a singularity at $k=n+1$ and this prevents the left expression in the large brackets from being evaluated. We can therefore choose to sum up to $n-1$ instead (then the boundary evaluation will occur at $k=n$ rather than $k=n+1$). With this,~\eqref{eq:ct} turns into
    \begin{equation}\label{eq:ctexample}
    \sum\limits_{k=5}^{n-1}(S_n-2)\binom{n}{k}-\underbrace{\left(\frac{n^4 - 6n^3 + 11n^2 - 30n}{24}\right)}_{\text{inhomogeneous part}}=0.
    \end{equation}
    This fixes the issue of the singularity (alternatively, we could have rewritten $\frac{k}{k-n-1}\binom{n}{k}=\frac{-k}{n+1}\binom{n+1}{k}$ to get rid of the pole). Next, we note that the upper summation bound contains the parameter $n$ while our telescoper $P$ contains the shift operator $S_n$: applying the operator to the whole sum affects both the upper bound and $\binom{n}{k}$. This is fixed with the (human) observation that
    \[
    \sum\limits_{k=5}^{n-1}(S_n-2)\binom{n}{k}=(S_n-2)\sum\limits_{k=5}^{n}\binom{n}{k}\underbrace{-\binom{n+1}{n}-\binom{n+1}{n+1}+2\binom{n}{n}}_{\text{compensated terms $=-n$}}.
    \]
    In this situation, we say that the operator and the summation does not commute. Then~\eqref{eq:ctexample} simplifies to the inhomogeneous recurrence
    \[
    (S_n-2)\sum\limits_{k=5}^{n}\binom{n}{k}=\frac{n^4-6n^3+11n^2-6n}{24}.
    \]
    If one prefers a homogeneous recurrence, an annihilating operator for the right-hand side can be determined to be $(n-3)S_n-(n+1)$. We can therefore conclude that
    \[
      \big((n-3)S_n-(n+1)\big)\cdot(S_n-2)=(n-3)S_n^2+(5-3n)S_n+2(n+1)
    \]
    is the desired annihilating operator for~\eqref{eq:binomialsum}. As is expected in such simple examples, the recurrence can be solved (by the computer) to obtain a closed form for~\eqref{eq:binomialsum}. Of course, it agrees with the one that one directly gets from invoking the binomial theorem.
    
    In the above example, we can see that there is a ``dance'' between human and the computer and only upon a careful collaboration does it bear fruit. We now proceed to use a similar strategy to attack the big sum $G_s(x)$ and furthermore present some alternatives to improve performance. The total computation time largely depends on how complicated the summands and inhomogeneous parts turn out to be after (human) simplification. The next section outlines some of these strategies and in particular highlights how we were able to successfully derive (and prove) a recurrence for $G_s(x)$. 
    
\section{A Playbook for the Holonomic Approach}

    This section illustrates how to generally overcome the difficulties listed in the previous section and how to effectively perform the human-computer dance to prove our main result. We envision that the discussion leads to a deeper understanding of the practical issues when applying the holonomic systems approach and makes it accessible for other applications. The Mathematica notebook containing implementations of these strategies can be found in the online supplementary material \cite{KoutschanWong20}.
    
    \begin{thm}\label{thm:main}
        For $b,m,s\in\N, b\geq 2$, the polynomial given in \eqref{eq:mainpoly},
        \[
          G_s(x):=\sum_{k=1}^{m+s-1}\Bigg(\sum_{r=1}^{s}\binom{s}{r}\binom{k-1}{r-1}\frac{b-1}{(-b)^r}\sum_{i=0}^{r-1-c_m(k)}(-b)^i\binom{r-1}{i}\Bigg)(bx)^k,
        \]
        with~$c_m(k):=\max(k-m,0)$, satisfies the recurrence
        \begin{align*}
            &(s+2)(b x-1)\cdot G_{\!s+3}\\
            &+\left(m(bx-1)(x-1)+bsx(x-2)+bx(x-3)-s(2x-3)-3 x+5\right)\cdot G_{\!s+2}\\
            &-(x-1)(b m x+b s x+b x+m x-2 m+s x-3 s+x-4)\cdot G_{\!s+1}\\
            &+(x-1)^2 (m+s+1)\cdot G_{\!s} = 0.
        \end{align*}
    \end{thm}
    We note again that this result is already contained in \cite[Lemma 15]{WiartWong20} with computational details found in~\cite{KoutschanWong20}. We also remark that a recurrence in~$m$ could also be derived, but for the application in question the above recurrence in~$s$ was sufficient. The following discussion serves to outline alternate (and in some cases, faster) proof strategies, to provide some exposition for technical details that were not mentioned in \cite{WiartWong20}, and to explicitly resolve some of the issues mentioned in the previous section in as much generality as possible under the context of using our problem as a case study. We hope that this will be useful for future practitioners.
    
    \subsection{Preprocessing the triple sum $G_s(x)$}
    
    Before we dive in, we make a few remarks about how to view $G_s(x)$ to make our life easier.  On the one hand, the summation~\eqref{eq:mainpoly} can be separated into two parts $G_s=G_s^{(1)}+G_s^{(2)}$, in order to remove the max function in the upper limit of the innermost sum.
    After a mild simplification (human), these two parts look as follows:
    \begin{align*}
        G_s^{(1)} & :=-\!\!\!\sum_{k=1}^{m+s-1}\sum_{r=1}^s\binom{s}{r}\binom{k-1}{r-1}\left(\frac{b-1}{b}\right)^r(bx)^k,\\
        G_s^{(2)} & :=\sum_{k=m+1}^{m+s-1}\,\sum_{r=1}^{s}\binom{s}{r}\binom{k-1}{r-1}\frac{1-b}{(-b)^r}\sum_{i=r-(k-m)}^{r-1}(-b)^i\binom{r-1}{i}(bx)^k.
    \end{align*}
    Observe that $-G_s^{(2)}$ is the collection of terms that is added to $G_s$ to enable the sum to collapse to $G_s^{(1)}$. We write this out to show that initially applying the full strategy to ``only'' the double sum $G_s^{(1)}$ gives a hypothetical lower bound for the time and effort required to treat the whole sum $G_s$. We note that if $k>m+s-1$, then there is no reason to expect that either of the inner sums would be zero, which may cause the inhomogeneous parts in~\eqref{eq:ct} to survive.

    The split sums also serve as an example of how to apply closure properties: the sum of holonomic functions is still holonomic \cite[Proposition 3.1]{Zeilberger90}, so 
    \[
      \ann\left(G_s^{(1)}+G_s^{(2)}\right)
    \]
    can be deduced by executing (computer) the corresponding ``closure property of addition'' algorithm after separately computing a respective annihilating ideal for each of the two terms. This closure property can also be applied in intermediate computations (for example, during the treatment of the inhomogeneous parts). However, the user should be aware that there is a risk that the recurrence order (more precisely: the holonomic rank) may increase during each such application (but not more than the sum of their orders). We learned that what we initially thought was a clever idea of splitting the sums, turned out to be less than optimal in terms of computational resources for this particular problem, in ways that will be described in the next few sections.

    On the other hand, we can also choose to deal with the full triple sum right from the start.
    By observing that when $k-m<0,$ we have the situation where $r-1<i\leq r-1-(k-m)$ forces the innermost binomial coefficient $\binom{r-1}{i}$ to be 0. Thus, the max function in $G_s(x)$ can be safely removed. We can also move all summations to the front and consider only one summand with three indexed parameters. In other words, $G_s(x)$ can be rewritten as
    \begin{equation}\label{eq:triplesum}
    \underbrace{\sum_{k=1}^{m+s-1}\sum_{r=1}^{s}\sum_{i=0}^{r-1-(k-m)}}_{\text{quantifiers grouped}}
    \underbrace{\vphantom{\sum_{k=1}^{m-1}}
    \binom{s}{r}\binom{k-1}{r-1}\binom{r-1}{i}\frac{b-1}{(-b)^{r-i}}(bx)^k}_{\text{one \vphantom{gqp}summand}}.
    \end{equation}

    In general, the creative telescoping algorithm works very well on these kinds of symbolic sums (in all forms as described above), and can be applied directly without adjustments if all conditions are ``ideal''. In such cases, the outputted telescoper corresponds exactly to the desired recurrence. Unfortunately, this kind of naive ``hey let's give it a try'' multiple/parallel application of creative telescoping to both the split sum case and to \eqref{eq:triplesum} resulted in some incorrect first-order recurrence (which was easily debunked by plugging in a few values). Hence, (human) adjustment was needed. For the split sum case, these adjustments produced extraneous terms that were not a part of the original sum and/or came from compensations due to the rebuilding of the original sum. We subsequently collected all such terms to find a collective annihilator for them for the purpose of ``homogenizing'' the recurrence given by the telescoper. All of these techniques are described in Sections \ref{sec:sing}--\ref{sec:sub}. For \eqref{eq:triplesum}, we used different approaches,
    and these are outlined in Sections~\ref{sec:gamma} and~\ref{sec:gf}.
    
    \subsection{Guessing}
    
    Another ``preparation'' step involves employing the Guess package~\cite{Kauers09} to predict the recurrence that our polynomial satisfies, by using sufficiently generic evaluations of~\eqref{eq:triplesum}, that is, we evaluate our polynomial for a finite number of values of our main variables $x,b,m,s$ and use the resulting data to reconstruct the coefficients of the recurrence with the command \textsc{GuessMultRE}. We furthermore impose what we believe is the general shape of the recurrence and if such a recurrence exists, the guessing procedure will produce one. This serves as an additional sanity check for future calculations: if the creative telescoping algorithm produced a recurrence of higher order, then we would know that it overshoots and we could try to find a different approach that would produce a better result. For our problem, the (computer) guessing procedure already produced the claimed minimal third-order recurrence (from Theorem \ref{thm:main}) in the parameter~$s$. This means that we know $G_s$ satisfies the recurrence for at least a finite number of values of~$s$. To prove that the guess is correct (i.e., that it satisfies the recurrence for an infinite number of~$s$), it is enough to compute the same recurrence (or a higher order one) via creative telescoping. In the latter case, one also has to verify that the guessed recurrence (operator) is a right factor of the bigger one, and consider a sufficient number of initial values. 
    
    \subsection{Dealing with Singularities}\label{sec:sing}
    
    We found that some certificates~$Q$ contain singularities on the boundary values of the inner sum. This implies that the limits of the sum must be adjusted so that we avoid evaluating at those points. In fact, the summation range must be adjusted so that there are no singularities at \textbf{all} intermediate values.
        
    To illustrate this a little more generally, suppose that, upon applying the creative telescoping algorithm on the summation $\sum_{r=1}^s F(s,r)$ with respect to~$r$, the computer outputs a certificate $Q \in \mathbb{Q}(s,r)$ containing poles at $r_i\in~[1,s+1]\subset\mathbb{N}$ for a finite number of $i$. All parameters besides $r$ are treated symbolically. Then we can see that the sum $\sum_{r=1}^s (S_r-1)\cdot Q \cdot F(s,r)$ cannot be determined because evaluations at those poles for the summation range $[1,s]$ are not possible and therefore the sum is undefined.
    
    Such singularities can be removed from the offending sum so that the evaluation(s) can happen. We also remove the exact same values from the summations containing the telescopers to match so that \eqref{eq:ct} makes sense. The summations with the telescopers can then be subsequently ``filled in'' with the removed summands (balanced of course by subtracting those same terms from the inhomogeneous part). This strategy can be quite effective if all of the poles are collected contiguously at either of the summation bounds, or if there are only one or two. In the case of the inner sum of $G^{(1)}_s$,  \[\sum\limits_{r=1}^s\underbrace{\binom{s}{r}\binom{k-1}{r-1}\left(\frac{b-1}{b}\right)^r(bx)^k}_{\text{summand}},\] applying creative telescoping produces a certificate $Q=\frac{-bxr(r-1)(k-s)}{(k-r+1)(r-(s+1))}$ corresponding to the telescoper $P=bsxS_s-(k+1)S_k+x(k-s)$. Clearly, $Q$ has singularity at $r=s+1$, which prevents us from certifying that $P$ is an annihilator for this sum. Similar to the toy example, the fix here is simply to acknowledge the relation \eqref{eq:ct} only up to $s-1$:
    \begin{equation}\label{eq:ctrelsing1}
    \sum\limits_{r=1}^{s-1} P\cdot \text{summand}-\sum\limits_{r=1}^{s-1}(S_r-1)\cdot Q \cdot \text{summand} = 0.
    \end{equation}
    The game is now to rearrange terms so that we can get the annihilator for $\sum_{r=1}^s \text{summand}$. The right summation in \eqref{eq:ctrelsing1} is a telescoping sum that can now be evaluated easily at the boundary terms. We refer to this as the ``delta part''. The remaining inhomogeneous terms will come from the left sum of~\eqref{eq:ctrelsing1}, where we insert the $r=s$ term and compensate for the insertion with $-P\cdot\text{summand}|_{r=s}$, which we will call the ``compensated term''. Thus, \eqref{eq:ctrelsing1} becomes
    \begin{equation}\label{eq:ctrelsing2}
    \sum\limits_{r=1}^{s} P\cdot \text{summand}+\underbrace{\left(\text{compensated term}+\text{delta part}\right)}_{\text{inhomogeneous terms}}=0.
    \end{equation}

    We are closer to our goal, but run into a new problem that $P$ contains a shift in $s$, which occurs in the upper limit. So, we cannot factor out this $P$ without a little more work. We will address this in the next section. Note that if the number of compensated terms required is too large, it may be better to consider another strategy, such as rewriting terms in some alternative (but equivalent) form to avoid the poles entirely (as was suggested in the analysis for~\eqref{eq:binomialsum}). But in general, there may not be an easy way rewrite such expressions: the ease in which this is possible depends on the properties of the objects at hand.
        
    \subsection{Pulling Operators Outside of the Sum}\label{sec:commute}
    
    We also found that the telescoper~$P$ does not commute with our summation. To illustrate this a little more generally, suppose the operator~$P$ is in the Ore algebra generated by the shift operator~$S_s$. In other words, suppose that $P$ can be written as some polynomial in $S_s$, for example, 
    \[
      P=p_0+p_1S_s+\cdots+p_jS_s^j,
    \]
    where the $p_i$ ($i=0,\ldots,j$) may be rational functions containing the parameter~$s$. If we apply such a $P$ to a summation of the form $\sum_{k=1}^{m+s-1}H(s,k)$, then we face the issue that the application not only affects the parameter $s$ in the summand $H(s,k)$, but also the upper limit $m+s-1$. Then if we apply $P$ to the whole sum, we get
    \begin{equation}\label{eq:commute1}
          p_0\!\!\sum\limits_{k=1}^{m+s-1}\!\!H(s,k) \;+\;
          p_1\!\sum\limits_{k=1}^{m+s}\!H(s+1,k) \;+\cdots+\;
          p_j\!\!\!\!\!\sum\limits_{k=1}^{m+s+j-1}\!\!\!\!\!H(s+j,k).
    \end{equation}
    It is quite obvious that this is not the same as applying $P$ to only the summand $H(s,k)$:
    \begin{equation}\label{eq:commute2}
        \sum\limits_{k=1}^{m+s-1} \Bigl(p_0H(s,k)+p_1H(s+1,k)+\cdots +p_jH(s+j,k)\Bigr).
    \end{equation}
        
    However, we can simulate the ``factoring out'' of the $P$ in~\eqref{eq:commute2} if we peel off a sufficient (and finite) number of terms from each sum in~\eqref{eq:commute1} such that its upper limits are all $m+s-1$. Then~\eqref{eq:commute2} can be replaced by the peeled version of~\eqref{eq:commute1} with $P$ on the outside and the removed terms can then be merged with the inhomogeneous part. 
    
    Continuing with our example from the previous section, we can see that we are a little lucky in that the $P$ in \eqref{eq:ctrelsing2} only has one shift in $s$ (the ideal case being that $P$ has no shift in~$s$). So in this case we just have to peel off one term $P\cdot\text{summand}|_{r=s+1}$, which we will call the ``comp $S$-shift''. Thus, \eqref{eq:ctrelsing2} becomes
    \[P\cdot \sum\limits_{r=1}^{s} \text{summand}+\underbrace{\left(\text{comp $S$-shift}+\text{compensated term}+\text{delta part}\right)}_{\text{inhomogeneous terms}}=0.\]
    Let $R$ be the annihilator for the inhomogeneous terms. Then $R\cdot P$ is the annihilator for $\sum_{r=1}^s\text{summand}$. Sometimes, $R$ is easy to compute, sometimes it is not. The next section addresses how to deal with $R$ if it is not.
    
    \subsection{Treatment of Inhomogeneous Parts}
    \label{sec:inhom}
        
    In all of our examples, the adjustment of the summation limits to avoid singularities in the certificates was completed first. After that, it is a~priori not clear if one should proceed to adjust for the operator commuting with the summand or to fill in the terms that were removed for the singularities. The (human) decision may depend on the number of singularities (bounded above by the degree of the denominators of the certificates), where the singularities are located, how complicated the telescoper expression is, and whether or not the lower/upper boundaries are influenced by the telescopers. This makes it difficult to automate adjustments effectively.
    
    Once we have collected all of the inhomogeneous parts, we face the question of how to process them. In principle, we could just write them down and try to use Mathematica's symbolic power to simplify them as much as possible. Unfortunately, this does not work very well on our problem, with the only progress being that some of the inhomogeneous parts conveniently collapse to zero (so we remove them). Instead, we take advantage of the fact that we can write all of the inhomogeneous parts as different shifts and substitutions of the given summand. More precisely, the total of some of these parts can be expressed as an operator applied to the summand, followed by a substitution. Then, an annihilator for this total can be derived by applying the closure properties ``application of an operator'' and later follwed by an ``integer-linear substitution''. In this way, we completely avoid dealing with expressions like Mathematica's DifferenceRoots.

    We illustrate a hands-on approach of finding an operator to apply with an example that comes from the inhomogeneous parts for $G^{(1)}$, which contain hypergeometric $_2F_1$ series that we write out in full detail below. The strategy involves observing patterns in a complicated expression to construct an operator that would give the same result when it is applied to some simpler version of the expression. Given
        \begin{align}\label{eq:inhomG1}
            \begin{split}
            \frac{(b-1)(m+s+1)(bx)^{m+s+1}}{b^2x}\cdot
            {}_2F_1\left(1-s,-m-s;2;\frac{b-1}{b}\right)\\
            -\frac{(b-1)(m+bs)(bx)^{m+s}}{b^2}\cdot
            {}_2F_1\left(1-s,1-m-s;2;\frac{b-1}{b}\right)\\
            +\frac{(b-1)(s+1)(bx-1)(bx)^{m+s}}{b}\cdot
            {}_2F_1\left(1-m-s,-s;2;\frac{b-1}{b}\right),
            \end{split}
        \end{align}
        we can see that selecting the operator
        \begin{align}\label{eq:inhomG1newop}
            \frac{(b-1)(m+s+1)}{b^2x(bx)}S_m^2-\frac{(b-1)(m+bs)}{b^2(bx)}S_m+\frac{(b-1)(s+1)(bx-1)}{b(bx)}S_s
        \end{align}
        and applying it to $(bx)^{m+s}\cdot {}_2F_1\bigl(1-s,2-m-s;2;\frac{b-1}{b}\bigr)$ results in~\eqref{eq:inhomG1}. Such an operator is certainly not unique and there is no hard rule to construct one, but with a little experimentation and a focused goal of attaining one reasonably ranked operator to be applied to one function (rather than three different ones), it can be deduced in a practical amount of time if such an operator exists. We can also make minor adjustments such as moving the factor $b-1$ from the operator to the function in the example above and deduce an annihilating operator in this way. However, minor changes like these will not bring about much improvement in computational efficiency. Thus, we want to emphasize that this part of the game is not really algorithmic, but rather part of the human's job to decide what is the most elegant.
        
        The annihilator for \eqref{eq:inhomG1} can therefore be obtained by ``applying''~\eqref{eq:inhomG1newop} (in the sense of closure properties) to the annihilator of this single expression. This process is much faster than trying to directly compute an annihilating ideal of the sum of hypergeometric series, with the added benefit that the order of the recurrence will usually be smaller compared to applying the \textsc{Annihilator} command directly to expressions such as~\eqref{eq:inhomG1}. In our particular example, the latter method even caused the program to crash.
                
        We can therefore see that directly constructing an operator by closely inspecting patterns in the inhomogeneous parts can be computationally effective. In a way, the fact that we use $_2F_1$'s in the previous argument is inconsequential to the construction of the operator to be applied (it could have easily be replaced with a symbolic expression that exhibits similar shift behaviors, for example). Thus, an annihilating operator for the inhomogeneous terms can be deduced in this way \textit{before} administering any substitutions. As mentioned before, this could also involve removing all terms that would collapse to zero anyway and building from scratch the new operator by only using the shifts needed to produce compensation terms that may have resulted from the treatment of singularities and commutation.
        
        Applying this strategy to the inhomogeneous parts of $G_s^{(2)}$, we get an acting operator that lies in the Ore algebra $\mathbb{Q}(b,k,m,s,x)(S_k,S_s)$ and has the support
        \begin{equation}\label{eq:G2R2act}
        \lbrace S_k^2S_s^3,S_kS_s^3,S_kS_s^2,S_s^3,S_s^2,S_k,S_s,1
        \rbrace.
        \end{equation}
        Comparing this with \eqref{eq:inhomG1newop} gives an indication why the annihilator computation for $G^{(2)}_s$ would take longer. Furthermore, after ``applying'' this new operator to the annihilator of the summand, we still had the additional step of making a substitution $k\rightarrow m+s$, costing us nearly 30 hours. 
        See the table in Figure \ref{fig:intermediaryresults} for a summary of the shapes and sizes of these objects. It is clear then, that while our problem benefited from this strategy, it is still not optimal and we proceed to present other ways to improve computational efficiency.
    
    \subsection{Substitution Speedup}\label{sec:sub}
    
    The collection of all inhomogeneous parts and its subsequent removal via its annihilator can eat up a lot of computation time depending on how complicated these parts are. In particular, we experience this in the computation of the annihilators for the inhomogeneous part of $G_s^{(2)}$ when applying the closure property of integer-linear substitution. Thus, as an alternative to blindly applying the corresponding computer command and not knowing what is going on behind the scenes while waiting patiently for the code to finish, we can take better control of the process by making a few additional optimizations, resulting in a significant speedup of computation time.
    
    With the ``application of an operator'' closure property using the method as described in the previous section we were able to produce an annihilating ideal, with its Gr\"{o}bner basis denoted by $U^{(2)}$ in the Ore algebra $\mathbb{Z}(b,k,m,s,x)(S_k,S_s)$, for the combined inhomogeneous parts of $G_s^{(2)}$, denoted by $H(s,k)$, but without the necessary substitution $k\rightarrow m+s$ according to the upper summation bound. This implies that it is necessary to apply the closure property ``integer-linear substitution'' to~$U^{(2)}$.
    
    It turns out that the above Gr\"obner basis $U^{(2)}$ has the set of irreducible monomials $\{S_s,S_k,1\}$ and hence is of holonomic rank~$3$. The theory tells us that for $H(s,m+s)$ we should expect a recurrence of order~$3$ in~$s$, however, we find by trial-and-error that an order~$2$ operator already works (which is in agreement with the result of the longer computation in Section~\ref{sec:inhom}).
      
    For constructing such a recurrence, we want to find an operator~$T$ in the left ideal generated by $U^{(2)}$ with the support $\{ S_s^2 S_k^2, S_s S_k, 1 \}$, corresponding to a bivariate recurrence involving the terms
    \[
      H(s+2,k+2),\; H(s+1,k+1),\; H(s,k),
    \]
    which, after the substitution $k\rightarrow m+s$, turns into the desired second-order recurrence for $H(s,m+s)$. We therefore make the following ansatz for~$T$:
    \[
      T=c_2(k,s)S_k^2S_s^2+c_1(k,s)S_kS_s+c_0(k,s),
    \]
    where the coefficients $c_0(k,s),c_1(k,s),c_2(k,s)$ are to be determined. Gr\"obner basis theory tells us that this~$T$ is an element of the annihilating ideal (in other words: represents a valid recurrence for $H(s,k)$) if and only if it reduces to~$0$ by the Gr\"obner basis~$U^{(2)}$. Reducing the ansatz~$T$ by $U^{(2)}$ results in a linear combination of the basis monomials $\{S_s,S_k,1\}$, i.e., an Ore polynomial of the form
    \[
      E_2(k,s,c_0,c_1,c_2)S_s + E_1(k,s,c_0,c_1,c_2)S_k + E_0(k,s,c_0,c_1,c_2)
    \]
    with rational functions $E_0,E_1,E_2$. This polynomial is zero if and only if $E_0=E_1=E_2=0$, so we proceed to solve this system for $c_0, c_1, c_2$. 
    
    Ultimately, this procedure gives us an Ore polynomial with the support we want
    (if we had chosen the support of $T$ too small, we would have realized that by getting no solution of the system $E_0=E_1=E_2=0$). This operator~$T$ 
    annihilates the inhomogeneous parts after substituting $k\rightarrow m+s$ in its coefficients and omitting the shift operator $S_k$. This is essentially the same as saying: substitute $k\rightarrow m+s$ into the recurrence $T\cdot H(s,k)$. However, since this substitution tends to decrease the size of expressions (it reduces the number of variables), it is desirable to perform it as early as possible, and not only at the very end.
    
    Indeed, we were able to speed up our computation significantly by performing the substitution already during the reduction of the monomials of~$T$, but care has to be taken: to match the leading monomials, we may have to multiply by (a power of) $S_k$, and for this (noncommutative) multiplication one needs to keep the variable~$k$. However, it can be substituted immediately afterwards. This leads to a less dramatic swell of expressions. Other sources of speedup include a manual selection strategy of Gr\"obner basis elements to be used for the reduction, and the order in which these reductions are made.

    It might be worthwhile to note that there are places in this process where we got lucky: as fate would have it, the coefficient $E_2$ of $S_s$ is zero (but only for $k=m+s$), and this gives us the luck of finding a recurrence of order two instead of three! This procedure has now reduced our total computation time from 30 (using the strategy of the previous section) to 1.4 hours.
    
    \begin{figure}[ht]
    \begin{center}
    \setlength{\tabcolsep}{0.1cm}
    \renewcommand{\arraystretch}{1.2}
    \resizebox{\textwidth}{!}{%
    \begin{tabular}{|l|l|l|l|l|l|l|}
        \hline
        \rule{0pt}{10pt}Sum& Method & Object & Time (s) & Rank & Shape &
        Bytes\\[2pt]
        \hline
        \multirow{8}{*}{$G^{(1)}_s$}& \multirow{2}{*}{CT} & \multirow{2}{*}{I1}&\multirow{2}{*}{fast}&\multirow{2}{*}{2}&$\lbrace S_k,S_s,1\rbrace,\lbrace S_s^2,S_s,1\rbrace$ & \multirow{2}{*}{3720}\\
        & & & & & $(1,1,0,1),(0,1,0,1)$ & \\\cline{2-7}
        & \multirow{2}{*}{CT} & \multirow{2}{*}{P1}&\multirow{2}{*}{fast}&\multirow{2}{*}{1}&$\lbrace S_s,1\rbrace$ & \multirow{2}{*}{1368}\\
        & & & & & $(1,1,0,0)$ & \\\cline{2-7}
        & \multirow{2}{*}{3.3--3.5} & \multirow{2}{*}{R1}&\multirow{2}{*}{fast}&\multirow{2}{*}{3}&$\lbrace S_s^3,S_s^2,S_s,1\rbrace$ & \multirow{2}{*}{339736}\\
        & & & & & $(6,3,6,6)$ & \\\cline{2-7}
        & \multirow{2}{*}{R1$**$P1} & \multirow{2}{*}{ann1}&\multirow{2}{*}{fast}&\multirow{2}{*}{4}&$\lbrace S_s^4,S_s^3,S_s^2,S_s,1\rbrace$ & \multirow{2}{*}{645528}\\
        & & & & & $(7,4,6,6)$ & \\
        \hline
        \multirow{12}{*}{$G^{(2)}_s$}&\multirow{2}{*}{CT}& \multirow{2}{*}{I2}&\multirow{2}{*}{7}&\multirow{2}{*}{3}&$\lbrace S_s^2,S_k,S_s,1\rbrace,\lbrace S_k,S_s,S_k,1\rbrace,\lbrace S_k^2,S_k,S_s,1\rbrace$ & \multirow{2}{*}{15720}\\
        & & & & & $(1,1,1,3),(1,0,0,1),(2,1,1,2)$ & \\\cline{2-7}
        & \multirow{2}{*}{CT} & \multirow{2}{*}{P2}&\multirow{2}{*}{70}&\multirow{2}{*}{3}&$\lbrace S_s^3,S_s^2,S_s,1\rbrace$ & \multirow{2}{*}{5120}\\
        & & & & & $(1,1,0,0)$ & \\\cline{2-7}
        & \multirow{2}{*}{3.3--3.5}&
        \multirow{2}{*}{R2act}&\multirow{2}{*}{330}&\multirow{2}{*}{3}&$\lbrace S_s^2,S_k,S_s,1\rbrace,\lbrace S_k,S_s,S_k,1\rbrace,\lbrace S_k^2,S_k,S_s,1\rbrace$ & \multirow{2}{*}{50918792}\\
        & & & & & $(10,5,5,11,12),(10,4,4,9,10),(11,5,5,11,12)$ & \\\cline{2-7}
        & \multirow{2}{*}{\ref{sec:inhom}} & \multirow{2}{*}{R2 (slow sub)}&\multirow{2}{*}{107870}&\multirow{2}{*}{2}&$\lbrace S_s^2,S_s,1\rbrace$ & \multirow{2}{*}{953768}\\
        & & & & & $(8,4,9,9)$ & \\\cline{2-7}
        & \multirow{2}{*}{\ref{sec:sub}} & \multirow{2}{*}{R2 (fast sub)}&\multirow{2}{*}{4200}&\multirow{2}{*}{2}&$\lbrace S_s^2,S_s,1\rbrace$ & \multirow{2}{*}{953768}\\
        & & & & & $(8,4,9,9)$ & \\\cline{2-7}
        & \multirow{2}{*}{R2$**$P2}& \multirow{2}{*}{ann2}&\multirow{2}{*}{fast}&\multirow{2}{*}{5}&$\lbrace S_s^5,S_s^4,S_s^3,S_s^2,S_s,1\rbrace$ & \multirow{2}{*}{3931560}\\
        & & & & & $(10,5,10,10)$ & \\
        \hline
        \multirow{2}{*}{$G_s(x)$} & \multirow{2}{*}{ann1+ann2} & \multirow{2}{*}{ann} & \multirow{2}{*}{16} & \multirow{2}{*}{5} & $\lbrace S_s^5,S_s^4,S_s^3,S_s^2,S_s,1\rbrace$ & \multirow{2}{*}{3931848}\\
        & & & & & $(10,5,10,10)$ & \\
        \hline
    \end{tabular}}
    \end{center}
    \caption{Some intermediary results for the split sum case. ``CT'' refers to the direct application of creative telescoping prior to any modifications, ``I'' refers to the annihilator of all of the inner sums, ``P'' refers to the telescoper resulting from CT applied to I, ``R2act'' refers to the result of the (human) constructed operator \eqref{eq:G2R2act} for the inhomogeneous parts acting on I2, ``R'' refers to the collective annihilator of the inhomogeneous parts (achieved by making an appropriate substitution into R2act), ``R$**$P'' indicates non-commutative multiplication to obtain the annihilator of the whole sum, ``ann'' is obtained by applying the closure property of addition by computing \textsc{DFinitePlus}[ann1,ann2], ``fast'' indicates computations of $<1$ second, ``Rank'' refers to the holonomic rank of the object, ``Shape'' provides information about the support (top) of the generators for the annihilating ideal and their coefficient degrees (bottom) in the variable order $(x,b,m,s)$ (and order $(x,b,m,s,k)$ for R2act).}
    \label{fig:intermediaryresults}
    \end{figure}

    \subsection{Gamma Insertions}\label{sec:gamma}

    In this section, we focus on dealing with the triple sum \eqref{eq:triplesum}. We remark that the naive approach (applying creative telescoping directly three times, sequentially or in one parallel step, but without any of the required adjustments) produces an incorrect first-order recurrence for $G_s(x)$, namely 
    \[(1-bx) S_s + (x-1)\] (plugging in a few values will confirm its incorrectness). However, this computation was quick. Unfortunately, the certificates
    \[\frac{b k - i k + b i k - b r + i r - b i r}{b^2 (i - r) (1 + r)},\ \frac{r(bx-1)}{r-(s+1)},\ \frac{i k - i r + b^2 i x - b i k x + b^2 i r x}{
    b^2 (i - r) (1 + r)},\]
    have the issue with the singularity at $r=s+1$. Moreover, since the telescoper contains a shift in $s$, the problem in dealing with the commutation still persists. Thus, the nice looking first-order recurrence is a little bit deceiving.
    
    This next strategy differs from the previous sections in that it takes a more holistic approach by treating \eqref{eq:triplesum} all at once and makes adjustments to the single summand in order to deal with the ``unnatural boundary'' problems simultaneously with the introduction of a new parameter. This idea can be dated back to the thesis of Wegschaider \cite[2, Section 2.7.3]{Wegschaider97} with the caveat that an extra parameter would increase computation time. For our problem however, we did not observe a significant increase in this regard.
    
\begin{figure}[ht]
    \centering
    \includegraphics[scale=0.4]{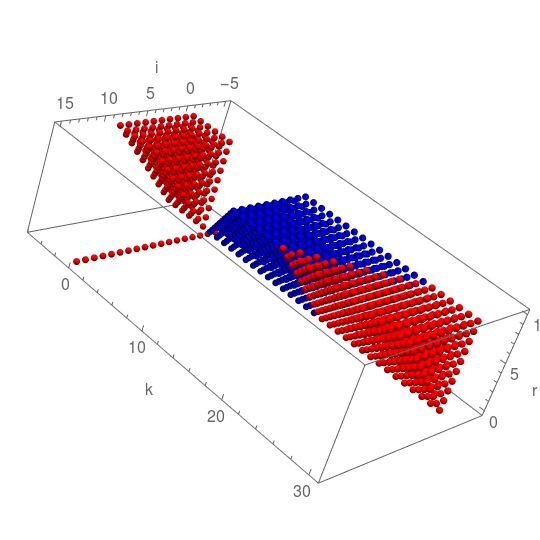}
    \caption{``Bad'' points $B_1\setminus B_2$ (red) and ``good'' points $B_1\cap B_2$ (blue) for fixed $m=15$ and $s=10$.}
    \label{fig:unbounded}
    \end{figure}

    The issue of ``unnatural boundaries'' occurs whenever all three binomial coefficients in our summand are nonzero beyond the limits of our summation. Assuming that $m$ and $s$ are fixed positive integers, we let $B_1$ denote the collection of points $(k,i,r)\in\Z^3$ for which the summand in~\eqref{eq:triplesum} is nonzero, and let $B_2$ denote all points $(k,i,r)\in\Z^3$ that are inside the summation ranges. While $B_2$ corresponds to all integer points of a bounded polytope in $\R^3$, the set $B_1$ is unbounded (see Fig.~\ref{fig:unbounded}). We essentially want to sum over all points $(k,i,r)$ in the intersection of $B_1$ and $B_2$ (depicted in blue), while we want to avoid those points in $B_1\setminus B_2$ (depicted in red). Hence, the set of ``bad points'' also forms an infinite polytope and we remove these points using gamma functions.
   
   We first recall that $\Gamma(k)$ has poles exactly at the non-positive integers, and therefore $\frac{1}{\Gamma(k)}$ has zeros at $k=0,-1,-2,\ldots$. Then the summand can be modified by the following gamma functions in order to enforce natural boundaries:
    \[
      C(\varepsilon,i,k,r):=\binom{s}{r}\cdot\binom{k-1}{r-1}\cdot\binom{r-1}{i}\cdot
      \frac{\Gamma(k+\varepsilon)}{\Gamma(k)}\cdot
      \frac{\Gamma(r-i-(k-m)+\varepsilon)}{\Gamma(r-i-(k-m))}
    \]
    with some new symbol~$\varepsilon$. Upon sending $\varepsilon\to0$ we get back the original product of binomial coefficients, while the two additional factors force the expression $C(\varepsilon,i,k,r)$ to be zero whenever $k\leq0$ or $r-i-(k-m)\leq0$. In other words, the introduction of the reciprocal gammas is balanced out by the perturbed gammas in the numerators, which conveniently avoids division by zero and gives an equivalent (final) result for the original problem after setting $\varepsilon=0$. We comment that the gamma function with the perturbation is a hypergeometric (and therefore holonomic) term that takes on a finite value away from the pole.

    We have now achieved a summand such that creative telescoping can be applied without having to worry about undesirable terms from beyond the summation boundaries. Another consequence of this is that the telescoper can be pulled out of the summation. Unfortunately, this procedure does not completely remove the threat of singularities that may show up in the certificates, so inhomogeneous parts can surface here and must be treated (this turned out to be the case in our situation). Except for that, the net effect of introducing gammas is so that we can apply the creative telescoping algorithm (three times) to the triple sum
    \[
      \sum_{k=1}^{m+s-1}\sum_{r=1}^s\sum_{i=0}^{r-1-(k-m)}C(\varepsilon,i,k,r)\cdot \frac{b-1}{(-b)^{r-i}}(bx)^k
    \]
    and afterwards take the limit $\varepsilon\to0$ to obtain the desired recurrence for $G_s(x)$. Unfortunately, we do not get the minimal-order recurrence, but a fourth-order one. This has allowed us to reduce our computation time to about 11 minutes.

    However, this is not yet the end of the story. In Fig.~\ref{fig:unbounded} we were trapped by Mathematica's definition of the binomial coefficient (cf. the discussion in Section~\ref{sec:background}): actually, the summand is zero for $k\leq0$ if we employ the intended ``correct'' definition, which implies that $\binom{n}{k}$ is zero unless $0\leq k\leq n$. The conditions for the summand to be nonzero (implied by the three binomial coefficients) somehow correspond to the summation bounds (given by the three summation quantifiers), which is illustrated in the following table (it is actually a curiosity of this problem that we can find such correspondence):
    \medskip
    
    \begin{center}
    \renewcommand{\arraystretch}{1.2}
    \setlength{\tabcolsep}{0.2cm}
    \begin{tabular}{|c|c|c|}
        \hline
        \multirow{2}{1.6cm}{Factor in Summand} & Nonzero Range & Summation Bounds\\
        & ($B_1$) & ($B_2$) \\
        \hline
        \rule[-7pt]{0pt}{19pt}$\binom{s}{r}$ & $0\leq r\leq s$ & $1\leq r \leq s$\\
        \hline
        \rule[-7pt]{0pt}{19pt}$\binom{k-1}{r-1}$ & $0\leq r-1\leq k-1$ & $1\leq k \leq m+s-1$\\
        \hline
        \rule[-7pt]{0pt}{19pt}$\binom{r-1}{i}$ & $0\leq i\leq r-1$ & $0\leq i \leq r-1-(k-m)$\\
        \hline
    \end{tabular}
    \end{center}
    \medskip

    After close inspection of this table, it becomes evident that only one gamma correction is actually needed (and hence Fig.~\ref{fig:unbounded} does not show the true situation). We can therefore redefine
    \[
      C(\varepsilon,i,k,r):=\binom{s}{r}\cdot\binom{k-1}{r-1}\cdot\binom{r-1}{i}\cdot \frac{\Gamma(r-i-(k-m)+\varepsilon)}{\Gamma(r-i-(k-m))}.
    \]
    This observation speeds up the computations significantly, and the winning time is 30 seconds. Moreover, we obtain the minimal recurrence of order three. The reader may now wonder how we can tell the HolonomicFunctions package that this computation should be executed with a different definition of the binomial coefficient (that differs from Mathematica's)? The answer is: we do not have to, since it is completely irrelevant (from the viewpoint of the package), because both versions of the binomial coefficient satisfy the very same recurrence equations, as we have seen in Section~\ref{sec:background}! The difference only becomes relevant when we evaluate the summand at particular values (which is done outside of the package), e.g., when checking initial conditions.
    
    \begin{figure}
    \begin{center}
    \setlength{\tabcolsep}{0.2cm}
    \begin{tabular}{|c|l|c|}
        \hline
        \rule{0pt}{10pt}Approximate & \multicolumn{1}{|c|}{Strategies} & \multirow{2}{*}{Result}\\
        Comp. Time & \multicolumn{1}{|c|}{Implemented} & \\[2pt]
        \hline
        \multirow{4}{*}{30 hours} & & \multirow{9}{4cm}{fifth-order recurrence in the ideal generated by the guessed recurrence} \\[-10pt]
        & -- split sums\rule{0pt}{10pt} $G^{(1)}_s$ and $G^{(2)}_s$ & \\
        & -- sing./comm. corrections  & \\
        & -- closure properties & \\[2pt]
        \cline{1-2}
        \multirow{5}{*}{1.4 hours} &&\\[-10pt]
        & -- split sums\rule{0pt}{10pt} $G^{(1)}_s$ and $G^{(2)}_s$ & \\
        & -- sing./comm. corrections  & \\
        & -- closure properties & \\
        & -- substitution speedup & \\[2pt]
        \hline
        \multirow{3}{*}{11 minutes}
        & -- the triple sum\rule{0pt}{10pt} \eqref{eq:triplesum}
        & \multirow{3}{4cm}{fourth-order recurrence in the ideal generated by the guessed recurrence} \\
        & -- two gamma insertions & \\
        & -- sing. corrections & \\[2pt]
        \hline
        \multirow{3}{*}{30 seconds}
        & -- the triple sum\rule{0pt}{10pt} \eqref{eq:triplesum}
        & \multirow{3}{4cm}{same third-order recurrence as the guessed recurrence} \\
        & -- one gamma insertion & \\
        & -- sing. corrections & \\[2pt]
        \hline
        \multirow{2}{*}{$<$ 1 second}
        & -- the triple sum \eqref{eq:triplesum}
        & \multirow{2}{4cm}{a generating function with $G_s(x)$ coefficients}\\
        & -- residues \cite{BostanLairezSalvy17} & \\[2pt]
        \hline
    \end{tabular}
    \end{center}
    \caption{Results and Comparisons}\label{fig:results}
    \end{figure}
    
    \subsection{Generating Functions of Binomial Sums}\label{sec:gf}
     
    In the final section of this chapter, we would like to highlight another symbolic computation approach to deal with binomial sums. It has a different flavor from creative telescoping but nevertheless allows us to also solve our problem in a most elegant way. In 2015, Bostan, Lairez, and Salvy studied the representation of the generating functions of binomial sums via integrals of rational functions \cite{BostanLairezSalvy17}. In their paper, they acknowledged the critical role that the certificate plays in the creative telescoping method and some of the computational problems that it creates. Thus, they sought to use complex analysis to provide an alternate way to view binomial sums without the need to deal with certificates. However, one would still need to take care of unnatural boundaries before applying their method. Fortunately, their implementation allows us to do this by accepting Heaviside functions in the input, which is analogous to our usage of gamma quotients with $\varepsilon$ in Section \ref{sec:gamma}. This means that by representing our triple sum as
    \[
    \sum_{k=1}^{\infty}\sum_{r=1}^{s}\sum_{i=0}^{\infty}\binom{s}{r}\binom{k-1}{r-1}\binom{r-1}{i}\frac{b-1}{(-b)^{r-i}}(bx)^k\cdot H(r-1+(k-m)-i),
    \]
    we can apply their package, and it returns the rational function
    \[
    \frac{xyz(b-1)}{(z-1)(y-1)(1-(1-x)z-yxb)}
    \]
    for which the generating function
    \[
    \sum\limits_{m=0}^{\infty}\sum\limits_{s=0}^{\infty} G_s(x)y^mz^s
    \]
    is its residue, with $z$ indexing the parameter~$s$ and $y$ indexing the parameter~$m$. In this way, the non-positivity of the coefficients (our triple sum) can be read off directly under the original conditions of $b>1$ and $x\in[0,1)$. Their method also has some limitations (the input class of admissible expressions) which we will not go into detail here, but it is clear that on the problem at hand it works extremely well: the whole computation to derive the above rational function took less than a second!
    
\section{Conclusions and Future Work}

In this expository article, we demonstrated the usage of the HolonomicFunctions package to deal with an intricate triple sum coming from an application in quasi-Monte Carlo integration. We had a couple of objectives in mind for this paper: first, we felt the need to deliver some technical details for a key lemma in~\cite{WiartWong20}, where only the main ideas of the computer algebra proof were mentioned; second, we wanted to provide a somewhat easy-to-digest description for proving special function and combinatorial identities with the help of the computer by expounding on the difficulties that may arise in similar applications and highlighting a few creative ways to cure them. We hope that we have convinced the reader that it is not always so cut-and-dry to prove a given identity with the holonomic systems approach. While in principle it allows one to prove holonomic identities in an automated way, we have seen that in practice, even with using current state-of-the-art software tools, many steps in the proof require human interaction. At many positions in the proving process, we had to make a choice on how to proceed, and the decision may both influence the optimality of the final result and the time that is required to obtain it. Fig.~\ref{fig:results} gives an impressive overview how much difference in runtime such choices can make.

It is also clear that some of these strategies have bits and pieces in existing literature. Some were already mentioned throughout the text, and we summarize here. For example, the idea of splitting sums to simplify computation problems was discussed by Prodinger \cite{Prodinger96}, Wegschaider briefly proposed the addition of a parameter for singularity removal in his thesis \cite{Wegschaider97} (another application of this can be found in \cite{LyonsPauleRiese02}), some of the details of treating boundary terms in multisums appeared in \cite{AndrewsPauleSchneider05} and \cite{ChyzakMahboubiSibutPinoteTassi14}, and of course, we directly applied the generating function method from \cite{BostanLairezSalvy17}. Therefore, it is pertinent to remark that the contributed value of this paper is to present a new context with which to deal with these kinds of sums and to compile various strategies into a user-friendly format. It might also be worthwhile to note that despite these ``known'' issues dating back to the 1990s, not much progress has been made to automate them (and for good reason). This paper further serves the purpose of underlining that much work can still be done on this front!

With all that being said, a future plan is to automate some of the proof steps that had to be done ``by hand'' in this case study, for example, in the analysis of singularities in the certificate(s) and dealing with the issue of commutation.

\subsection*{Acknowledgements}
We would like to thank the organizers of CASC 2020 for providing an occasion and opportunity to give a talk about the work on which this article is based. We were encouraged by the positive feedback from the audience, which motivated this post-proceedings contribution. We would especially like to acknowledge Pierre Lairez for pointing us to his paper \cite{BostanLairezSalvy17} and for demonstrating how to make the computation in Section~\ref{sec:gf} with his binomial sums Maple package. We thank the reviewers for their careful reading, which helped us improve this manuscript greatly, particularly the second reviewer who pointed out some related literature and provided insightful criticism. We also express our appreciation to Hao Du and Ali Uncu for their support and helpful commentary.

%
%

\bibliography{main} 
\bibliographystyle{splncs04}

\end{document}